\documentclass[twocolumn]{aastex631}

\usepackage{amsmath}
\usepackage{breqn}

\shorttitle{Alfv\'en wave dynamics from crustal oscillations }
\shortauthors{Most, Kim, Chatziioannou\& Legred}
\graphicspath{{./}{figures/}}

\begin{document}

\title{Nonlinear Alfv\'en-Wave Dynamics and Premerger Emission from Crustal Oscillations  in Neutron Star Mergers}

\author[0000-0002-0491-1210]{Elias R. Most}
\email{emost@caltech.edu}
\affiliation{TAPIR, Mailcode 350-17, California Institute of Technology, Pasadena, CA 91125, USA}
\affiliation{Walter Burke Institute for Theoretical Physics, California Institute of Technology, Pasadena, CA 91125, USA}
\author[0000-0002-4305-6026]{Yoonsoo Kim}
\affiliation{TAPIR, Mailcode 350-17, California Institute of Technology, Pasadena, CA 91125, USA}
\affiliation{Department of Physics, California Institute of Technology, Pasadena, CA 91125, USA}

\author[0000-0002-5833-413X]{Katerina Chatziioannou}
\email{kchatziioannou@caltech.edu}
\affiliation{Department of Physics, California Institute of Technology, Pasadena, CA 91125, USA}

\affiliation{LIGO Laboratory, California Institute of Technology, Pasadena, CA 91125, USA}

\author[0000-0002-9523-9617]{Isaac Legred}
\affiliation{Department of Physics, California Institute of Technology, Pasadena, CA 91125, USA}
\affiliation{LIGO Laboratory, California Institute of Technology, Pasadena, CA 91125, USA}

\begin{abstract}

Neutron stars have solid crusts threaded by strong magnetic fields.
Perturbations in the crust can excite non-radial oscillations, which can in turn
launch Alfv\'en waves into the magnetosphere. 
In the case of a compact binary close to merger involving at least one neutron
star, this can happen through tidal interactions causing resonant excitations that shatter the neutron star crust. 
We present the first numerical study that elucidates the dynamics of Alfv\'en waves launched in a compact binary magnetosphere.
We seed a magnetic field perturbation on the neutron star crust, which we then evolve in fully general-relativistic force-free electrodynamics using a GPU-based implementation.
We show that Alfv\'en waves steepen nonlinearly before reaching the orbital light cylinder, form flares, and dissipate energy in a transient current sheet.
Our results predict radio and X-ray precursor emission from this process.
\end{abstract}


\section{Introduction}

Merging compact binaries are prime targets for gravitational-wave observations. 
In the case where one of the binary constituents is a neutron star the merger can be accompanied by electromagnetic counterparts, including afterglows and gamma-ray bursts~\citep{Cowperthwaite:2017dyu,Chornock:2017sdf,Villar:2017wcc,Nicholl:2017ahq,Troja:2018ruz,Tanvir:2017pws,Drout:2017ijr,LIGOScientific:2017zic,Savchenko:2017ffs,Troja:2017nqp,Margutti:2017cjl,Margutti:2018xqd,Hajela:2019mjy,Hallinan:2017woc,Alexander:2017aly,Ghirlanda:2018uyx,Mooley:2017enz,Mooley:2018qfh}. 
In addition, the presence of strong magnetic fields prior to the collision could also give rise to yet unseen precursor transients \citep{Hansen:2000am,Palenzuela:2013hu,Most:2022ayk,Cooper:2022slk}, which present exciting opportunities for next-generation observatories \citep{Corsi:2024vvr}.
For example, short-duration gamma ray bursts are associated with neutron star mergers~\citep{kumar2015physics,Gottlieb:2023sja,meszaros2006gamma} as confirmed by the GRB170817A burst coincident with a gravitational-wave event~\citep{LIGOScientific:2017zic}.
While the main burst is likely associated with the merger itself,
$\sim10\%$ of the observed long and short bursts are accompanied by precursor emission~\citep{Troja:2010zm,2020PhRvD.102j3014C,Xiao:2022quv,Dichiara:2023goh,2020PhRvD.102j3014C}. 
Preceding the main burst by $\sim 100\, \rm s$ to a few seconds for
long bursts ~\citep{2020PhRvD.102j3014C} and $\sim 2$s for short bursts \citep{Wang:2020vvr}, including a discovery of quasi-periodicity in a burst precursor accompanying a kilonova afterglow and a long gamma-ray burst~\citep{Xiao:2022quv}, precursor emission is associated with the binary inspiral.
One explanation links precursor emission to resonant excitation of modes in the neutron star crust~\citep{Tsang:2011ad,Penner:2011br,Tsang:2013mca,Neill:2021lat,Zhang:2022qtd} or in neutron star oceans in the inspiral~\citep{Sullivan:2022fsk,Sullivan:2023hxm}. 

In these models, binary tidal interactions excite oscillation modes in the crust-core interface of the neutron star(s) in a compact binary. 
These excitations cause the neutron star crust to fracture and release large amounts of energy, $>10^{46-47}\, \rm erg/s$~\citep{Tsang:2011ad,Penner:2011br}. 
For example, torsional modes could be consistent with the $20\,\rm Hz$ quasi-periodic oscillations claimed in GRB211211A~\citep{Suvorov:2022ldw}. 
Observation of such modes can put constraints on the equation of state of the crust, and in turn on the nuclear symmetry energy~\citep{Neill:2022psd,Sotani:2023lhu,Neill:2024gfe} or the presence of deconfined quark matter in the core~\citep{Pereira:2022apt}.
Such constraints would add to ground-based nuclear scattering experiments~\citep{Horowitz:2014bja,Reed:2021nqk,Reed:2023cap} and yield multi-messenger constraints on the equation of state complementary to those obtained through gravitational-wave~\citep{Raithel:2019uzi,Chatziioannou:2020pqz} or X-ray observations~\citep{Ozel:2016oaf,Watts:2016uzu}.

While estimates of the crustal energy budget and the tidal interactions are straightforward~\citep{1989ApJ...343..839B,Baiko:2018jax} (see also~\citet{Tsao:2021fdo,Sagert:2022gwu} for simulations of crustal oscillations), it remains unclear how waves launched from star quakes could convert into observable radiation.
Non-radial shearing modes will likely launch Alfv\'en waves. 
Such waves have been investigated in the context of gamma-ray bursts from neutron star quakes \citep{1989ApJ...343..839B,1993MNRAS.265L..13B}, as well as recently in the context of coincident Fast-Radio and X-ray bursts~\citep{Yuan:2020ayo,Yuan:2022uqt} from a galactic magnetar~\citep{CHIMEFRB:2020abu,Bochenek:2020zxn}.
In this picture,
Alfv\'en waves propagate from the neutron star surface into the magnetosphere, become nonlinear, and cause the emission of an ultra-relativistic blast wave (flare), as well as direct dissipation through magnetic reconnection~\citep{Yuan:2020ayo}.

Whether and how these dynamics can happen in a compact binary magnetosphere is the subject of this work. 
Numerical simulations of magnetospheric transients have matured considerably over the past years~\citep{Palenzuela:2010nf,Palenzuela:2012my,Palenzuela:2013kra,Alic:2012df,Ponce:2014hha,Nathanail:2020fkp,Carrasco:2020sxg,Carrasco:2021jja,East:2021spd,Most:2020ami,Most:2022ayk,Most:2022ojl,Most:2023unc}, including studies of the flaring dynamics in single~\citep{Parfrey:2013gza,Carrasco:2019aas,Mahlmann:2023ipm,Sharma:2023cxc} and binary neutron star magnetospheres~\citep{Most:2020ami,Most:2022ayk,Most:2023unc,Most:2022ojl}.
Leveraging this progress, in this work we perform magnetospheric simulations of crustal shattering flares and their nonlinear Alfv\'en-wave dynamics.

The aim of this paper is to clarify the magnetospheric dynamics of the crustal shattering scenario. 
In particular, we target the fraction of energy that can be converted into the final flare/blast wave state, and what the feedback of the flare on the magnetosphere might be, e.g., in the context of shock powered radio emission mechanisms~\citep{Beloborodov:2019wex,Metzger:2019una}.
To this end, we present global general-relativistic force-free electrodynamics simulations of Alfv\'en-wave dynamics in binary neutron star and black hole -- neutron star systems prior to merger. 
We demonstrate the nonlinear evolution of Alfv\'en waves, show the formation of blast waves at large distances, and place constraints on the energy conversion efficiency in flare launching and reconnection-mediated dissipation. 
Our results provide insight into the crustal shattering transient picture.

This work is structured as follows: We summarize the basic motivation of our study in Sec.~\ref{sec:picture}. We then describe the numerical setup and the binary configuration we study in Sec.~\ref{sec:methods}. 
Our main results and analyses are presented in Sec.~\ref{sec:results}, before concluding in Sec.~\ref{sec:conclusions}.

\section{Basic picture}
\label{sec:picture}

In the resonant crust shattering picture, tidal interactions between the neutron star and its companion excite modes at the crust-core interface prior to merger~\citep{Tsang:2011ad,Penner:2011br,Sullivan:2022fsk,Zhang:2022qtd}. 
The crust-core interface ($i-$)mode frequency is~\citep{Neill:2020szr}  
\begin{align}\label{eqn:f_imode}
    f_{i-\rm mode} \approx 130 - 170\, \rm Hz\,,
\end{align}
depending on the nuclear symmetry energy. 
Other modes have also been considered, including g-modes \citep{Passamonti:2020fur,Kuan:2021jmk,Kuan:2021sin}. For the i-mode,
the excitation timescale is~\citep{Tsang:2011ad,Neill:2020szr}
\begin{align}
    t_{i-\rm mode} \approx 0.04\, {\rm s}\, \left(\frac{1.2 M_\odot}{\mathcal{M}}\right)^{5/6} \left(\frac{150\, \rm Hz}{f_{i-\rm mode}}\right)^{11/6}\,,
\end{align}
where $\mathcal{M}$ is the binary chirp mass.
The excitation injects energy into the crust, which gets released once the crust fractures~\citep{Tsang:2013mca,Neill:2021lat} leading to estimated luminosities of \citep{Tsang:2013mca},
\begin{align}
     \mathcal{L}_{\rm crust} \simeq 10^{47}\, \left(\frac{B_{\rm NS}}{10^{13}\, \rm G}\right)^2\, \rm erg/s\,,
\end{align}
where $B_{\rm NS}$ is the neutron star 
surface magnetic field strength. 
One caveat is that the energy stored in the entire crust could be as low as~\citep{Baiko:2018jax}
\begin{align}
    E_{\rm crust} \gtrsim 2.5\times 10^{41} \left(\frac{r_{\rm NS}}{12\,\rm km}\right)^2 \,\rm erg\,,
\end{align}
where $r_{\rm NS}$ is the neutron star radius. 
Such energies would likely result in luminosities lower than those of gamma-ray burst precursors~\citep{2020PhRvD.102j3014C}.

\subsection{Launching Alfv\'en waves}

Crustal shattering will inject perturbations into the surrounding magnetosphere. 
These perturbations have been mainly associated with non-radial modes~\citep{1988ApJ...325..725M}, including either torsional~\citep{Suvorov:2022ldw}, or interface (i-)modes at the crust-core boundary~\citep{Tsang:2011ad}. 
Such non-radial modes will predominantly excite Alfv\'en waves along the magnetic field lines threading the crust. 
Modeling of a crustal quake has shown that the efficiency factor for converting the crustal energy release into Alfv\'en waves is $f\simeq 1\%$~\citep{2020ApJ...897..173B}, due to large energy absorption by the superfluid core. 
This reduces the overall luminosity injected into the magnetosphere,
\begin{align}
     \mathcal{L}_{\rm Alfv\acute{e}n}= f \mathcal{L}_{\rm crust} \simeq 10^{45}\, \left(\frac{f}{0.01}\right) \left(\frac{B_{\rm NS}}{10^{13}\, \rm G}\right)^2\, \rm erg/s\,.
\end{align}

\begin{figure}
    \centering
    \includegraphics[width=0.47\textwidth]{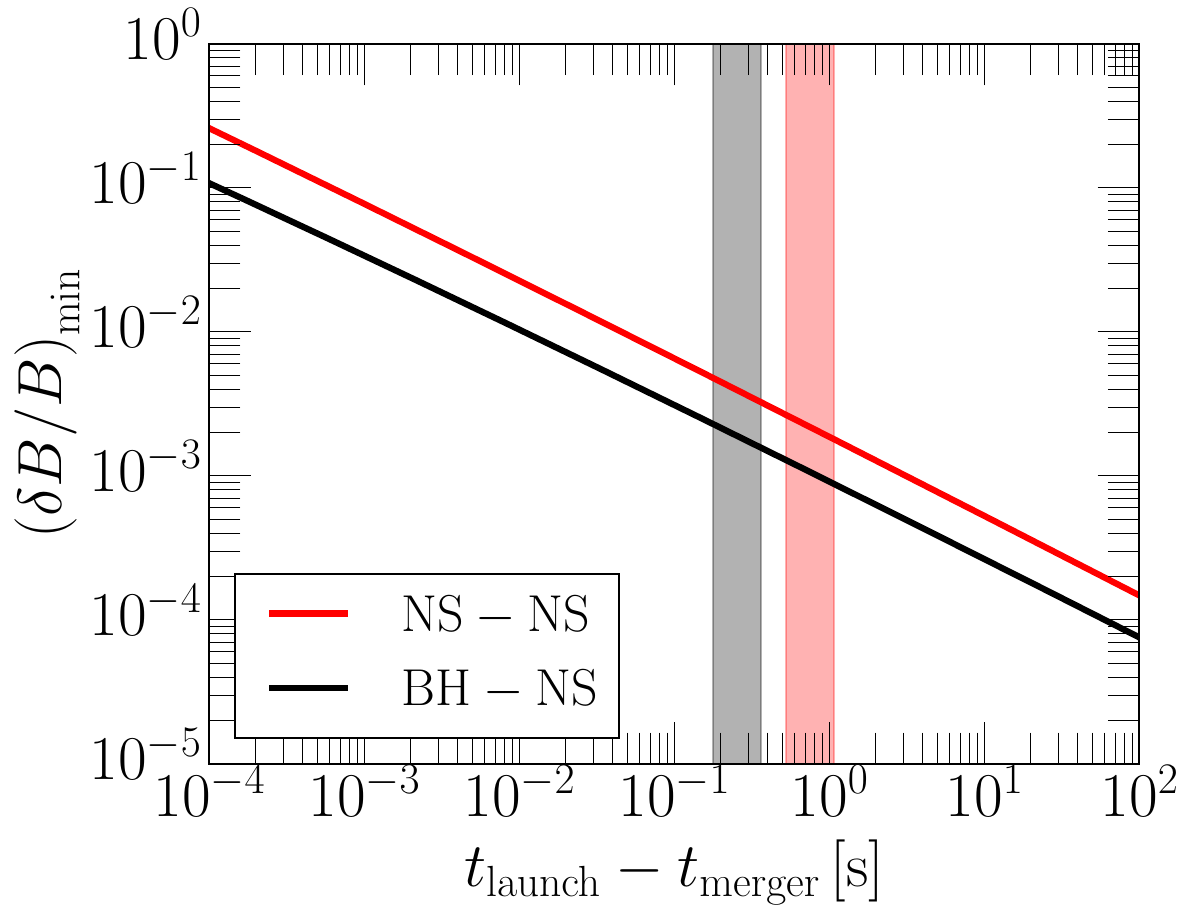}
    \caption{Minimum required perturbation that becomes
  nonlinear within the orbital light cylinder. The launching time, $t_{\rm launch}$, is given
  for black hole (BH) -- neutron star (NS) and binary neutron star (NS-NS)
  systems relative to the time of merger $t_{\rm merger}$. We adopt $m_2 = 1.4 \, M_\odot$, $r_{\rm NS} = 12\, \rm km$ and vary $Q=1$ (NS-NS) to $Q=4$ (BH-NS). Shaded regions correspond to the expected time window of the crust-core interface mode resonance from Eq.~\eqref{eqn:f_imode}.}
    \label{fig:dB_overB}
\end{figure}

Alfv\'en waves propagate as an initial perturbation along the magnetic field lines.
Following~\citet{1989ApJ...343..839B}, the resulting perturbation, $\delta B/B$, of the magnetic field, $B$, at the surface of the neutron star is
\begin{align}
    \left. \frac{\delta B}{B}\right|_{\rm surface} \simeq 0.02 \left(\frac{\mathcal{L}_{\rm Alfv\acute{e}n} }{10^{45}\, \rm erg/s}\right)^{1/2} \left(\frac{10^{13}\, \rm G}{B_{\rm NS}}\right)\,.
    \label{eqn:deltaB_min}
\end{align}
As a result of flux freezing and a magnetic dipole background, $B\propto r^{-3}$, decaying with radius, $r$, the perturbation steepens as
\begin{align}
    \frac{\delta B}{B} = \left(\frac{\delta B}{B}\right)_{\rm surface} \left(\frac{r}{r_{\rm NS}}\right)^{3/2}\,. \label{eqn:dB_overB}
\end{align}
The initial perturbation, $\left({\delta B}/{B}\right)_{\rm surface}$, will then grow as it propagates outwards, until it becomes nonlinear, approximately corresponding to $\delta B \simeq B$ \citep{1989ApJ...343..839B}. 
In case of multi-polar fields, $B\sim r^{-n}$ ($n>3$), non-linearity can be achieved earlier, and closer to the binary. At the same time, closed field lines can be shorter (compared to a dipole) placing tighter constraints on the required initial perturbation to reach non-linearity. The modified topology of the field would also affect the geometry of wave propagation. We will mainly discuss dipole fields in the following estimates.\\
Irrespective of the precise structure of the magnetic field, steepening will only efficiently happen on closed magnetic field lines which decay faster with radius than open ones, 
implying that nonlinear steepening needs to happen before the waves reach the orbital light cylinder,
\begin{align}
    r_{\rm LC} \simeq \frac{a^{3/2}}{\left(GM\right)^{1/2}}\,,
\end{align}
where $M=m_1+m_2$ is the binary total mass, $m_i$ are the component masses, $a$ is the orbital separation, and $G$
the gravitational constant.
We further define the mass ratio $Q = m_1/m_2 \geq 1$.
The largest radial distance an Alfv\'en wave can propagate inside the light cylinder is from the lighter companion $m_2$ in the direction of $m_1$.\footnote{For sufficiently large binary separation we can neglect the field lines interacting with/threading the companion as their volumetric fraction will be suppressed by $\propto \left(r_{\rm NS}/a\right)^3 \ll 1$.}
The lighter binary component is offset from the center of mass by $a Q/(1+Q)$, implying from Eq.~\eqref{eqn:dB_overB} that the smallest surface perturbation leading to nonlinear steeping is
\begin{align}\label{eqn:dB_over_B_min}
    \left(\frac{\delta B}{B}\right)_{\rm min} = \frac{{r_{\rm NS}}^{3/2} }{\left[ a\left( \sqrt{\frac{a}{GM}} + \frac{Q}{1+Q}\right)\right]^{3/2}} \approx \frac{{\left(GM\right)^{3/4}\, r_{\rm NS}}^{3/2} }{a^{9/4}} \,.
\end{align}
To leading order, the time to merger is \citep{Peters:1964zz}
\begin{align}\label{eqn:t_merger}
    t_{\rm merger} = \frac{5}{256} \frac{c^5}{G^3} \frac{a^4}{m_2^3 Q (1+Q)}\,,
\end{align}
leading to $\left(\delta B/B\right)_{\rm min} \propto t_{\rm merger}^{-9/16}$. Here $c$ is the speed of light. 
Equations~\eqref{eqn:dB_over_B_min} and~\eqref{eqn:t_merger} together yield the minimum perturbation required for the system to feature nonlinear flaring dynamics and are depicted in Fig.~\ref{fig:dB_overB}. 
Alfv\'en waves will non-linearly steepen for initial perturbations as low as $\delta B/B \sim 10^{-3}-10^{-4}$ if launched seconds before merger, which is less than the perturbation estimated in Eq.~\eqref{eqn:deltaB_min}. 
This back-of-the-envelop estimate implies that nonlinear Alfv\'en-wave dynamics is possible for crustal shattering induced by the i-mode resonance of Eq.~\eqref{eqn:f_imode} as indicated by the shaded regions in Fig.~\ref{fig:dB_overB}, see also~\citet{Tsang:2011ad,Neill:2021lat}).

\subsection{Observational signatures}

Assuming for the moment that any Alfv\'en wave will nonlinearly steepen before leaving the orbital light cylinder, the main question concerns its subsequent evolution. 
Will it produce a flare/relativistic blast wave that could potentially power electromagnetic transients? How are the flares formed, what energy do they carry?
While nonlinearity is commonly associated with the formation of charge-starved electric zones, numerical work suggests that the dissipation in charge-starved Alfv\'en waves is likely small~\citep{Chen:2020otx}.

Insights into potential emission mechanisms come from related simulations of star quakes on isolated neutron stars~\citep{Yuan:2020ayo,Yuan:2022uqt}. In these simulations, the Alfv\'en wave non linearly stretches, opens, and pinches off part of the closed magnetosphere. 
This leads to the ejection of a large plasmoid, which can be interpreted as a flare, similar to magnetar giant flares~\citep{Parfrey:2013gza} or binary neutron star flaring dynamics~\citep{Most:2020ami}. 
Consistent with~\citet{Yuan:2020ayo}, our simulations confirm this picture, Sec.~\ref{sec:results}. This is similar in spirit to flare collision models proposed for gamma-ray precursors from crustal shattering \citep{Tsang:2011ad,Neill:2021lat}. 
We further show that the conversion efficiency, $\eta<1$, from the Alfv\'en wave into the flare implies a final luminosity of
\begin{align}\label{eqn:Alfven_total}
    \mathcal{L}_{\rm flare} \simeq  2\cdot 10^{44}\, \left(\frac{\eta}{0.2}\right) \left(\frac{f}{0.01}\right) \left(\frac{B_{\rm NS}}{10^{13}\, \rm G}\right)^2\, \rm erg/s\,.
\end{align}
Furthermore, the flare has a trailing current sheet leading to direct dissipation with luminosities
\begin{align}
    \mathcal{L}_{\rm x-ray} \leq 0.1\, \mathcal{L}_{\rm Alfv\acute{e}n}\,,
\end{align}
in X-rays~\citep{Yuan:2020ayo,Yuan:2022uqt}, see also \citet{Most:2020ami,Most:2022ayk,Most:2023unc} for similar estimates in compact binary magnetospheres.

Collisionless shocks can produce coherent radio emission when a blast wave interacts with a surrounding wind~\citep{Metzger:2019una,Beloborodov:2019wex}. 
This process critically depends on the wind properties including the amount of baryon loading, as quantified by the wind magnetization $\sigma_w$.
Following \citet{Beloborodov:2019wex}, we estimate the emission peak frequency as 
\begin{align}
    \nu_{\rm peak} \simeq 5\,{\rm GHz} \left(\frac{10^{13}\, \rm cm}{r_s}\right)^{3/2}  \left(\frac{\Gamma_w}{20}\right) \left(\frac{E_{\rm flare}}{10^{41}\, \rm erg} \right)^{1/2}\,,
\end{align}
where $\Gamma_w$ is the wind Lorentz factor, and $r_s$ the distance from the stellar surface to shock formation.
Depending on the formation distance of the collisionless shock such a mechanism can produce radio emission.

One of the challenges of this model is that the wind density cannot be too low, or stated differently, the wind luminosity needs to be a fraction of the flare luminosity. 
For an orbiting dipole~\citep{Hansen:2000am,Ioka:2000yb}
\begin{align}
    \mathcal{L}_{\rm wind} &= \frac{4}{15c^5} \mu^2 a^2 \Omega^6\,,  \label{eqn:orbiting_dipole}\\
     &\approx 2.8\cdot 10^{39} \frac{{\rm erg}}{\rm s} 
    \left(\frac{B_{\rm NS}}{10^{13}\, \rm G}\right)^2 \left(\frac{M}{2.7\, M_\odot}\right)^3 
    \left(\frac{120\, \rm km}{a}\right)^7 \nonumber\,,
\end{align}
where $\Omega$ is the orbital angular speed, and $\mu$ the magnetic dipole moment.
In the context of multiple flares, the first flare could 
enhance the background in its upstream leading to $\mathcal{L}_{\rm wind} \sim 10^{-2} \, \mathcal{L}_{\rm flare}$~\citep{Yuan:2020ayo}. 
As we will demonstrate in Sec.~\ref{sec:results_flares}, this result also holds for binary magnetospheres.

The energy, $E_{\rm FRB}$, available to power a fast-radio burst is
\begin{align}
    E_{\rm FRB} \simeq 10^{38} {\rm erg}\, \left(\frac{\varepsilon_s}{10^{-3}}\right)\left(\frac{\mathcal{L}_{\rm flare}}{10^{44}\, \rm erg/s}\right) \left(\frac{t_{\rm burst}}{1\, \rm ms }\right)\,,
\end{align}
where $t_{\rm burst}$ is the observed burst duration, and $\varepsilon_s$ is the efficiency of the shock maser~\citep{Plotnikov:2019zuz,Sironi:2021wca}.
In the scenario where the previous flare enhances the wind (see discussion above), the burst duration is~\citep{Yuan:2020ayo,Beloborodov:2019wex}
\begin{align}
    t_{\rm burst} &= \frac{\Delta t_{\rm flare}}{4 \sigma_w} \left(\frac{\mathcal{L}_w}{\mathcal{L}_{\rm flare}}\right)^{1/2} \\
    &\simeq 1\, {\rm ms}\, \frac{1}{\sigma_w} \left(\frac{t_{i-{\rm mode}}}{40\, \rm ms}\right)  \left(\frac{\mathcal{L}_{\rm wind}}{10^{-2} \mathcal{L}_{\rm flare}}\right)^{1/2}\,,
\end{align}
where $\Delta t_{\rm flare} \sim t_{\rm i-mode}$ is the delay between subsequent flares. 
Here we have assumed that the crust can fracture multiple times on the resonant pumping timescale $t_{\rm i-mode}$.

\section{Methods} \label{sec:methods}

We numerically study the propagation and nonlinear dynamics of Alfv\'en waves in a compact binary magnetosphere, with an application to resonant shattering flares in neutron star mergers. 
This requires us to model the dynamics of the compact binary magnetosphere from the neutron star radius, $r_{\rm NS}$, where the perturbation is seeded, to $\simeq 50-100\, r_{\rm NS}\ \simeq 2-4\, r_{\rm LC}$, where the flaring dynamics and propagation happens. 
Bridging this large range of scales is not computationally feasible with kinetic models of magnetospheric plasmas. 
In addition, we are mainly interested in the global dynamics of the flaring process, as well as bulk estimates of energy conversion efficiencies in the system.

The simplest framework which includes Alfv\'en waves and can model the dynamics of the pair-plasma filled force-free magnetosphere~\citep{Goldreich:1969sb,Hansen:2000am,Lyutikov:2018nti} 
is force-free electrodynamics~\citep{Gruzinov:1999aza,Palenzuela:2010nf,Parfrey:2013gza,Carrasco:2019aas} in a full general-relativistic setting, e.g., \citep{Alic:2012df,Palenzuela:2012my,Paschalidis:2013jsa,Mahlmann:2020nwe,Most:2023unc,Kim:2024mau}.
Specifically, we numerically solve the Maxwell equations,
\begin{align}
    \nabla_\mu\!{^{\ast}\!F^{\mu\nu}} &= 0\,,\\
    \nabla_\mu  F^{\mu\nu} &= -\mathcal{J}^\nu\,,
\end{align}
where $\mathcal{J}^\nu$ is the electric current, and we have defined the Maxwell field strength tensor,
\begin{align}
F^{\mu\nu} = n^\mu E^\nu -  n^\nu E^\mu + \varepsilon^{\mu\nu\kappa} B_\kappa\,,
\end{align}
as well as its dual, $\!^{\ast}\!F^{\mu\nu}$, and electric, $E^\mu$, and magnetic, $B^\mu$, fields in the frame defined by a normal observer $n_\mu$.
The force-free conditions are enforced by adopting a resistive current~\citep{Alic:2012df,Palenzuela:2012my},
\begin{align}
    \mathcal{J}_i = q \varepsilon_{ijk} \frac{E^j B^k}{B^2} + \sigma \frac{E_j B^j}{B^2} B_i\,,
\end{align}
where $q$ is the electric charge density and $\sigma$ is the parallel conductivity, see, e.g.,~\citet{Palenzuela:2012my,Dionysopoulou:2012zv,Paschalidis:2013jsa,Palenzuela:2013hu,Palenzuela:2013kra,Ponce:2014hha,East:2021spd,Carrasco:2021jja} for applications in compact binary magnetospheres.

Following \citet{Most:2023unc} and \citet{Paschalidis:2013jsa}, we model the binary spacetime in general relativity using a fixed-orbit numerical solution of the extended conformally thin sandwich system~\citep{Grandclement:2006ht,Taniguchi:2007xm,Taniguchi:2007aq,Foucart:2008qt,Tacik:2016zal}.
The system is solved numerically using the \texttt{Kadath/FUKA}~\citep{Papenfort:2021hod,Grandclement:2009ju} code.
We carry out simulations of both binary neutron star and black hole -- neutron star systems. 
We assume that the stellar light cylinder lies outside the orbital light cylinder, i.e., the stars rotate slower than the orbital period~\citep{Bildsten:1992my, Zhu:2017znf}, in which case stellar spins will not affect the Alfv\'en wave dynamics. 
We, therefore, 
model the neutron star(s) as irrotational. This assumption is valid as long as the orbital frequency is larger than the stellar rotation frequency at the time of launching the initial perturbation. In the case of an i-mode resonance, Eq.~\eqref{eqn:f_imode}, this would hold except for millisecond pulsars. 
This implies that the only relevant intrinsic scales in the system (apart from the initial perturbation $\delta B/B$) are the orbital light cylinder, $r_{\rm LC}$, and to a lesser extent the neutron star radius, $r_{\rm NS}$, and the mass ratio $q$.  
System parameters are given in Tab.~\ref{tab:initial}.

\begin{table}
  \centering
\begin{tabular}{|l|c|c|c|c|c|}
    \hline
    \hline
    &$\Omega_{\rm orb}$ [$\rm s^{-1}$]  & $r_{\rm LC} [\rm km]$ & $a\, \left[\rm km\right]$ &
    $\chi_1$ & $\theta_{\rm B} [^\circ]$\\
    \hline
\hline
        BH-NS$^\ast\_0$ & 996.7 & 301&89.9 & -0.30  &   0\\
        BH-NS$^\ast\_30$ & 996.7&301 &89.9 & -0.30 & 30\\
        BH-NS$^\ast\_60$ & 996.7&301 &89.9 & -0.30 & 60\\
        BH-NS$^\ast\_90$ & 996.7&301 &89.9 & -0.30 & 90\\
        \hline
        NS-NS$^\ast\_0$ &  1021 &293&66.4 &  0.00 & 0\\
        NS-NS$^\ast\_30$ & 1021&293 &66.4 & 0.00&  30\\
        NS-NS$^\ast\_60$ & 1021&293 &66.4 & 0.00 & 60\\
        NS-NS$^\ast\_90$ & 1021&293 &66.4 & 0.00&  90\\
\hline
\hline
  \end{tabular}
	\caption{System parameters of the black hole (BH) -- neutron star (NS) and NS-NS binaries considered in this work. Columns mark 
        the fixed orbital frequency, $\Omega_{\rm orb}$, light cylinder, $r_{\rm LC}$, separation,  $a$, the dimensionless spin of the primary component, $\chi_{1}$, and the inclination angle, $\theta_{\rm B}$, of the magnetic dipole field of the secondary neutron star. For the BH-NS system we choose $m_{\rm 1/BH} = 5 \, M_\odot$ and $m_{\rm 2/NS} = 1.4\, M_\odot$, whereas the binary NS system is modelled as an equal mass system with total mass $M=2.7\, M_\odot$. All systems have $\chi_2=0$, and in the case of NS-NS systems, we set the primary magnetic field to be always anti-aligned with the orbital axis.}
    \label{tab:initial}
\end{table}

Simulations use the same numerical framework as in previous works simulating the dynamics of compact binary magnetospheres~\citep{Most:2020ami,Most:2022ojl,Most:2022ayk,Most:2023unc}.
We solve the general-relativistic force-free electrodynamics system using a relaxation approach with an effective parallel conductivity, which enforces the main force-free condition, $E_i B^i =0$~\citep{Alic:2012df,Palenzuela:2013kra}. 
Different from previous studies using this code, we implement a version of the ECHO scheme~\citep{DelZanna:2007pk} (see also \citet{Most:2019kfe}), combined with WENO-Z reconstruction~\citep{Borges2008}, and a Rusanov Riemann solver~\citep{Rusanov1961a}. 
The time integration for the stiff relaxation term is handled using the IMEX-SSP433
scheme~\citep{pareschi_2005_ier}, making the scheme formally third-order accurate in time, and fourth-order accurate in space. 
In addition, we manually enforce the second force-free constraint ($E^2<B^2)$ by rescaling $E^2 < 0.999 B^2$ where necessary.
This approach is consistent with the physically expected behavior at magnetospheric shock formation~\citep{Beloborodov:2023lxl}.
The numerical code is implemented on top of the
adaptive mesh-refinement (AMR) infrastructure of the \texttt{AMReX} framework~\citep{amrex}. 

\subsection{Computational challenges \& GPU computing}

The large numerical grid resolutions required for tracking the waves as they propagate from the neutron star surface to the orbital light cylinder and become nonlinear poses a computational challenge. 
Previous work on Alfv\'en-wave propagation in single-star magnetospheres has employed resolutions of $(8192\times 4096)$ in axisymmetry~\citep{Yuan:2020ayo}, and $(2560^3)$ in three dimensions~\citep{Yuan:2022uqt}. 
In this work, we have found it necessary to employ a total of 8 levels of mesh refinement, with a total number of more than $10^9$ grid points. 
The outer boundary extends to about $2,400\, \rm km$ in order to decouple reflections and spurious artefacts arising near the corners of the outer domain.

Given these computational challenges, we were forced to port our numerical code entirely to GPUs using the functionality of the \texttt{AMReX} infrastructure. 
With a performance of about $58\,$Mio cell-updates/s on a single A100 GPU, we have been able to carry out these simulations on $192$ A100 GPUs on the NERSC Perlmutter compute cluster. 
The total cost of the simulations shown here is about $10,000$ GPU node-hours.
To check the validity of our results, we have further run one simulation with substantially larger refinement regions on 2,400 V100 GPUs on OLCFs Summit system, which resulted in consistent dynamics.

\begin{figure}
    \centering
    \includegraphics[width=0.49\textwidth]{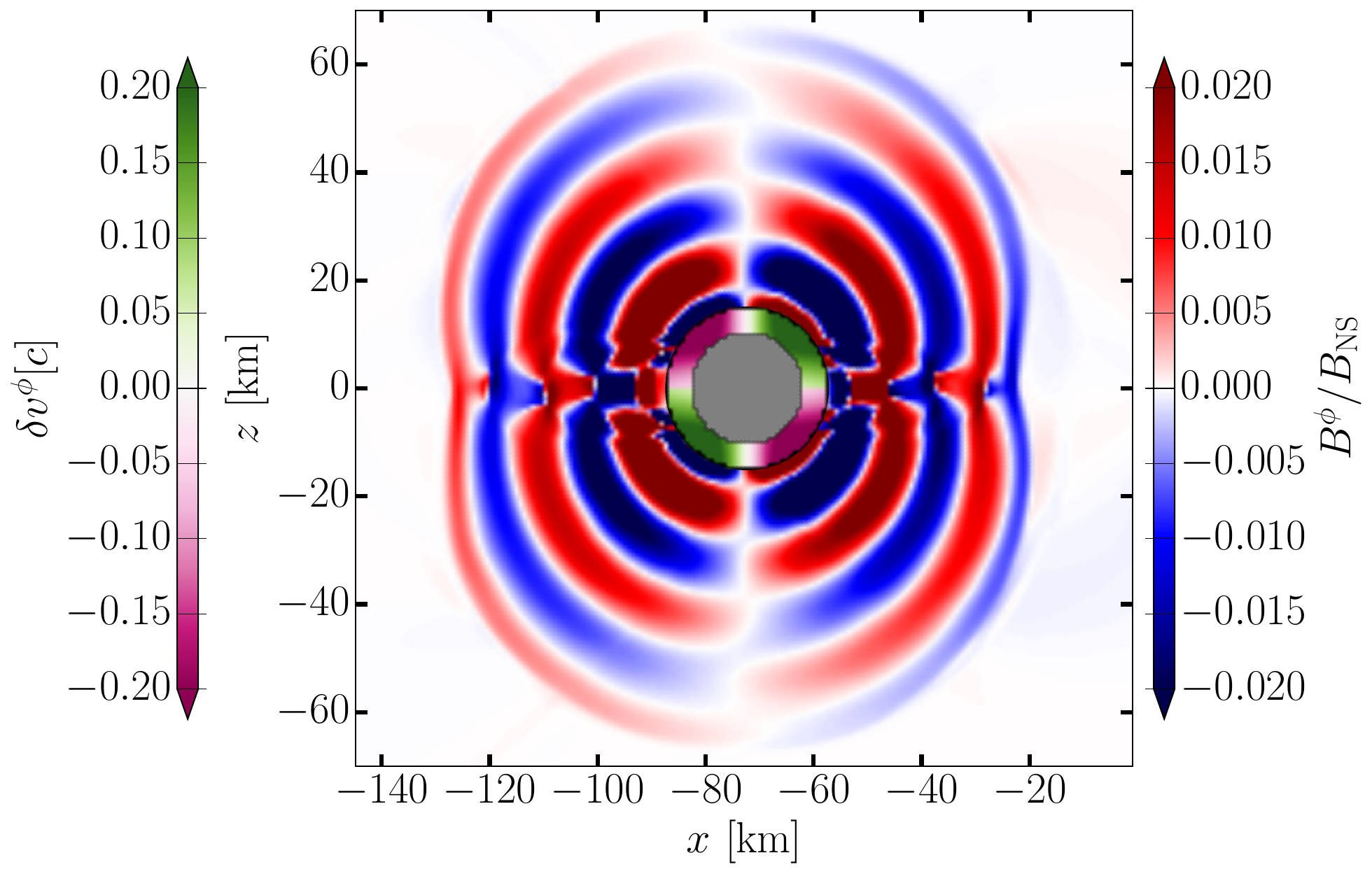}
    \caption{Initial wave packet of Alfv\'en waves launched from the neutron star (gray disk) for the aligned configuration, $\theta_B=0$.
    The orbital angular momentum points along the positive $z$-axis.
    Shown is the out-of-plane magnetic field component $B^\phi$, rescaled to the surface magnetic field strength $B_{\rm NS}$, as well as the relative surface perturbation velocity $\delta v^\phi$.}
    \label{fig:wave_id}
\end{figure}

\subsection{Wave launching}\label{sec:initial_wave}

We model the feedback of crustal shattering on the magnetosphere as a
wavetrain of monochromatic Alfv\'en waves with a fixed period and number of
wave cycles. For realistic crust shattering, likely complicated fracturing patterns \citep{Thompson:2016dkd} and resonances in the crust \citep{2020ApJ...897..173B} will inject a variety of different wavelengths.\\
Since the tidal interaction will mainly excite non-radial modes~\citep{Suvorov:2022ldw}, we model the surface motion following recent three-dimensional simulations of self-consistent neutron star oscillations. 
In particular, \citet{Sagert:2022gwu} reported that axisymmetric surface motion may resemble a bottle cap (polar) twist, see also \citet{Parfrey:2013gza}. Inspired by these results, we adopt
\begin{align}
    \delta \Omega\left(\theta\right) = \sigma A \exp\left({-\frac{\left({\theta-\theta_0}\right)^2}{2 \bar{\theta}^2}}\right) \cos\left(\omega \left(t-t_0\right)\right)\,,
\end{align}
where $\theta$ is the azimuthal angle in the coordinate of the neutron star, and we use $\left(\sigma=1, \theta_0=0\right)$ for the upper and $\left(\sigma=-1, \theta_0=\pi\right)$ for the lower hemisphere of the star. The resulting velocity perturbation $\delta v^\phi \sim \delta \Omega\, r_{\rm NS} \sin \theta$ is shown in Fig.~\ref{fig:wave_id}.
We inject this surface motion via an angular velocity at the outer layer of the neutron star surface, $\Omega_{\rm NS} = -\Omega_{\rm orb} + \delta \Omega$, for a total of four periods, after an initial time of $t_{\rm launch}$, chosen such that the magnetosphere can fully develop after the initial numerical transient has propagated away.
In the case of an equal-mass neutron star binary the perturbation is identical and sets in at the same time for both neutron stars. 
Since the wave will grow regardless of its initial value at the surface (see Sec.~\ref{sec:picture}), we fix a value of $\delta \Omega$ that comfortably leads to steepening of the wave within the computational domain we simulate. 
Specifically, we adopt a fixed perturbation of 
$\bar{\theta} = \pi/3$, $\omega M_\odot = 2 \pi/20$, and $A = 3 \pi \Omega_0$, which corresponds to $\delta B/B \sim 0.05$ at the surface of the neutron star.
We show the initial wave launching phase in Fig.~\ref{fig:wave_id}.

\section{Results}
\label{sec:results}

We investigate the nonlinear dynamics of Alfv\'en waves in
the magnetosphere of a compact binary in its final orbits prior to merger.
Specifically, we are interested in understanding the formation and dynamics
of flares caused by resonant shattering of the neutron star
crust due to gravitational tides in the inspiral~\citep{Tsang:2011ad,Penner:2011br}.
We present results as follows: first, we summarize
the state of the magnetosphere prior to launching Alfv\'en waves, Sec.~\ref{sec:results_intro}; second, we describe the magnetospheric dynamics of Alfv\'en waves
and the formation of flares, Sec.~\ref{sec:results_wave}, before presenting a parameter study on the neutron
star magnetic field, Sec.~\ref{sec:results_parameters}; finally, we summarize several properties of the flares, Sec.~\ref{sec:results_flares}.

\subsection{Background magnetosphere}\label{sec:results_intro}

We provide a summary of the background compact binary magnetosphere
prior to launching of Alfv\'en waves.
The magnetosphere consists of different regions.
Similar to a pulsar magnetosphere, the rotational motion of the orbit
creates a light cylinder at $r_{\rm LC} = c/\Omega_{\rm orb}$, dividing the
magnetosphere into zones of closed, $r < r_{\rm LC}$, and open, $r > r_{\rm LC}$, field lines. 
As discussed in Sec.~\ref{sec:picture}, it
is the closed zone where Alfv\'en waves will primarily steepen.
Unlike a pulsar magnetosphere~\citep{Spitkovsky:2006np,Kalapotharakos:2008zc}, the binary magnetosphere will in most cases not reach a quasi-stationary state~\citep{Palenzuela:2013kra,Most:2020ami,Carrasco:2021jja}.  

For neutron star
magnetic moments not aligned with the orbital axis, the magnetosphere will
feature periodic transient phases.\footnote{These transients are entirely due to the binary companion. Numerical~\citep{Carrasco:2020sxg} and
  analytical studies~\citep{Wada:2020kha} of single orbiting neutron stars show that global non-stationary phases will be entirely absent, although kinetic transients like gap discharges will still operate \citep{Bransgrove:2022afn}.}
In the case of black hole -- neutron star systems, magnetic field lines
from the neutron star will thread the black hole~\citep{Paschalidis:2013jsa,Most:2023unc}. These connected flux
tubes will lead to a built-up of a net twist due to orbital motion, ultimately
dissipating energy in a unipolar inductor-like scenario~\citep{1969ApJ...156...59G,Piro:2012rq,Lai:2012qe}. For small magnetic field inclinations, $\theta_B \lesssim 45^\circ$, the energy in the
twist will largely be dissipated in a trailing current sheet of the black
hole (see also magnetic draping \citep{Lyutikov:2022cuc}). For inclinations larger than this threshold, the twisted flux tubes will explode periodically
leading to flaring outbursts~\citep{Most:2023unc}. The
precise threshold for flaring may depend critically on the reconnection
rate in the black hole current sheet, which force-free simulations cannot
model correctly~\citep{Parfrey:2018dnc}. From a topological point
of view this should happen about twice per orbit~\citep{Cherkis:2021vto}.

For binary neutron star systems the situation is similar. For systems where
both stars carry a magnetic field with magnetic moments not fully aligned
in the same direction, connected flux tubes can be formed. Once a critical
twist is reached, these flux tubes will always reconnect and produce flares~\citep{Most:2022ojl}. Since both neutron stars will have crusts that can
shatter, this could lead to counter-propagating Alfv\'en-waves, which could interact and, in principle, drive turbulence~\citep{Goldreich:1994zz,TenBarge:2021qmk,Ripperda:2021pzt}.

\begin{figure*}
    \centering
\includegraphics[width=\textwidth]{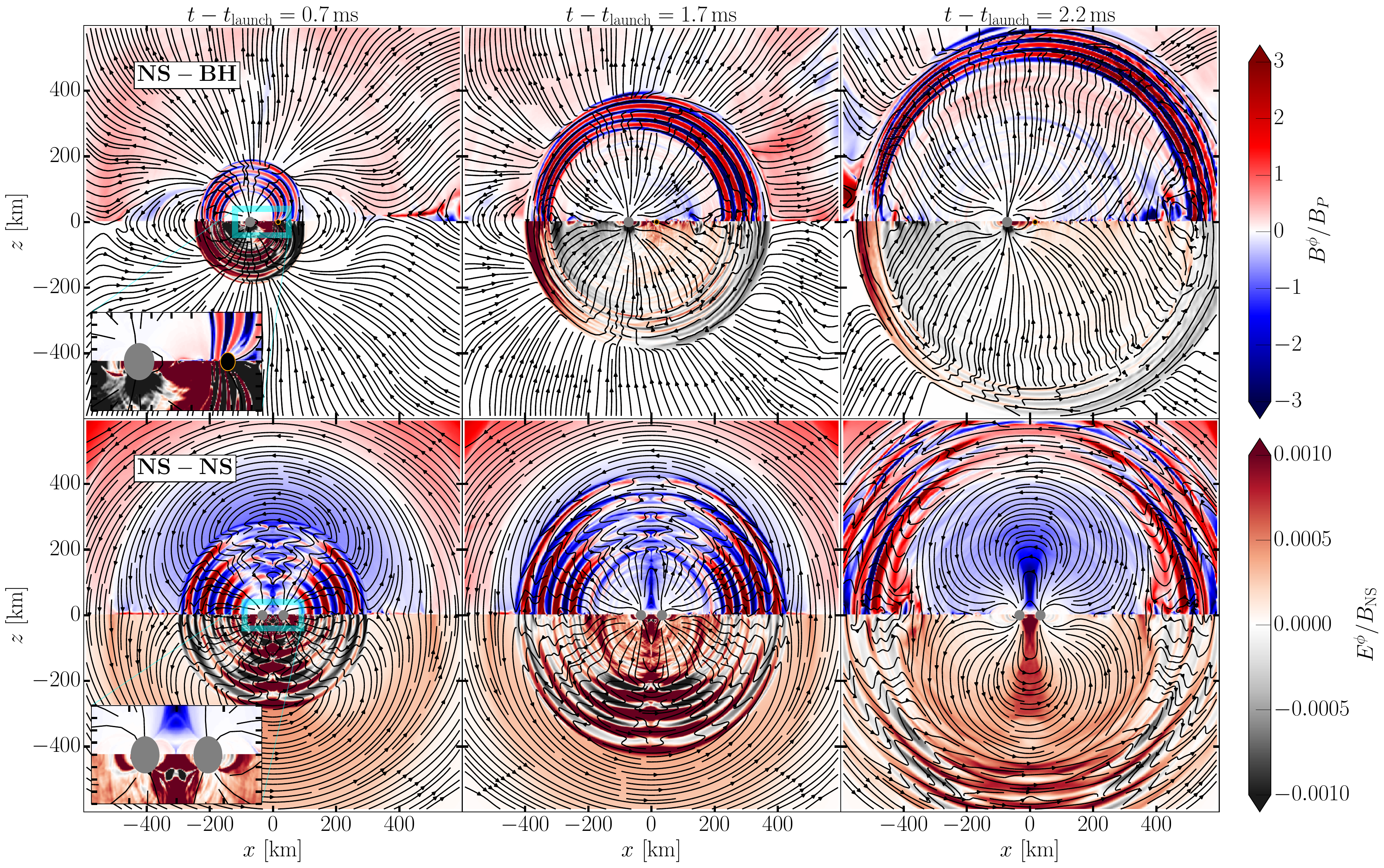}
    \caption{Alfv\'en wave propagation in a compact binary spacetime with aligned magnetic fields, $\theta_B=0^\circ$, in a frame co-rotating with the binary.
    (Top) Neutron star (NS) -- black hole (BH), (Bottom) NS-NS. In the NS-NS case, the companion NS has an anti-aligned magnetic field.
    Shown are three different times, $t$, stated relative to the time of launch $t_{\rm launch}$. Colors denote the out-of-plane magnetic, $B^{\phi}$, and electric, $E^{\phi}$, fields normalized to the poloidal magnetic field $B_{P}$ and the field strength at the surface $B_{\rm NS}$ respectively. 
    }
    \label{fig:NSBH_aligned}
\end{figure*}

\subsection{Alfv\'en wave launching and dynamics}\label{sec:results_wave}

Having established the background state of the binary magnetosphere, we now focus on the dynamics of Alfv\'en waves from their launch through their full nonlinear evolution. 
In principle, this process will not only produce Alfv\'en waves and especially should the fields contain any initial twist, the nonradial shearing motion from crust shattering will also produce fast waves~\citep{Mahlmann:2023ipm,Sharma:2023cxc}, which can steepen into monster shocks and also power transients \citep{Chen:2022yci,Beloborodov:2023lxl,Most:2024qgc}. However, given that the energetics will be a fraction of that of the Alfv\'en waves, we restrict to the latter.

The initial monochromatic wave package, see Sec.~\ref{sec:initial_wave}, is inject after a simulation time of $t_{\rm launch} \simeq 1,500\, {\rm km}/c$. 
At this time, the initial transient from the start of the simulation has fully propagated away from the inner regions, providing a clean background magnetosphere. 
As the wave package is propagating outward in the dipole magnetosphere of the neutron star it was launched in, the waves begin to steepen with radius following Eq.~\eqref{eqn:dB_overB}. 

Before discussing the simulation outcomes, we first briefly review general propagation properties of Alfv\'en waves.
Mathematically, Alfv\'en wave propagation is described by the relativistic Els\"asser equations~\citep{Chandran:2017zdg,Elsasser:1950zz} in the force-free limit,
\begin{align}
    \nabla_\nu \left(z_\pm^\mu z_\mp^\nu\right) &= - \left(\frac{3}{4} z_\pm^\mu z_\mp^\nu + \frac{1}{4} z_\mp^\mu z_\pm^\nu  + \frac{1}{2} g^{\mu\nu}\right)\frac{\partial_\nu b^2}{b^2}\,,\\
    z_\pm^\mu &= u_d^\mu \mp \frac{b^\mu}{b}\,,
\end{align} 
where $\left(u_d\right)_i = \gamma_d v_i$ with $\gamma_d = \sqrt{B^2/(B^2-E^2)}$ being the drift Lorentz factor, $v_i = \varepsilon_{ijk} E^j B^k / B^2$ being the drift velocity, and $b^\mu = B^\mu/\gamma_d$ the magnetic field in the drift frame. 
Alfv\'en waves propagate as perturbations, $\delta z_\pm^\mu = z_\pm^\mu - \left<z_\pm^\mu\right>$, relative to the background $\left<z_\pm^\mu\right>$.
Here, the sign in $z_\pm^\mu$ indicates the propagation of the wave relative to the magnetic field direction. Most importantly, the relativistic Els\"asser equations imply that for a magnetically dominated plasma \citep{TenBarge:2021qmk}, 
\begin{align}\label{eqn:d_elsasser}
    v_{A\mp}^\nu \partial_\nu \delta z_\pm^\mu = - \delta z_\mp^\nu \partial_\nu \delta z_\pm^\mu\,,
\end{align}
which is an advection equation with the Alfv\'en vector, $ v_{A\mp}^\nu$, with an interaction term depending on the orientation and propagation direction of the waves.
In particular, the interaction term will vanish if there are no counter-propagating Alfv\'en waves, which holds for any magnetization. 
Unlike for magnetosonic fast waves, this means that multiple Alfv\'en waves propagating in the same direction will not interact, even if they propagate at different speeds\footnote{In force-free electrodynamics, the Alfv\'en speed $v_A = \sqrt{\sigma/(1+\sigma)} \rightarrow 1$, as the magnetization $\sigma \rightarrow \infty$. While this is appropriate for magnetospheric dynamics, it also means that individual wave sections can never intersect.} and catch up with one another.

Equipped with this theoretical framework, we can now present and interpret our numerical results. 
We show the propagation of Alfv\'en waves in Fig.~\ref{fig:NSBH_aligned} for both black hole -- neutron star (mixed) binaries and binary neutron star systems. 
In the background dipole magnetosphere (shown in the co-orbiting frame) Alfv\'en waves will propagate as $B^\phi$ components of the magnetic field.

In the case of a mixed binary (top row), the waves can propagate in the direction of the black hole companion or away from it, which adds an additional redshift of the wavelength of the wave package as seen at infinity. 
Following discussion of Eq.~\eqref{eqn:d_elsasser}, we find no interaction of the Alfv\'en waves as expected from the absence of counter-propagating waves in the initial setup. 
As the waves propagate outward, an $E^\phi$ component of the electric field forms, which in this magnetic topology indicates the production of an outgoing flare, see the formation of a plasmoid (closed magnetic field line region) on the equator ~\citep{Yuan:2020eor,Bernardi:2024upq,Mahlmann:2024gui}.
Concurrently, the closed zone is ripped open by the nonlinear perturbation of the Alfv\'en waves, pinching off the outer part of the closed zone and ejecting a large plasmoid on the orbital plane (since the magnetic moment is aligned with the orbital axis). 
This demonstrates the launching of a flare due to Alfv\'en-wave nonlinearity. 
Trailing the flare is a current sheet, which will cause substantial additional dissipation of energy, potentially powering X-ray transients~\citep{Yuan:2020ayo}, see Sec.~\ref{sec:results_flares} for discussion.

The situation shown here for the mixed binary is very similar to the case of a star quake on a magnetar~\citep{Yuan:2020ayo,Yuan:2022uqt}. 
There, the quake launched an Alfv\'en wave localized to a specific flux bundle, which caused the emission of a flare within the hemisphere of the initial quake.
The crustal shattering case examined here with its large-scale perturbation is an extreme limit of that scenario, in which {\it all} field lines are excited. 
That said, the field lines on which fast waves can be produced are strongly constrained by the latitude of the perturbation relative to the magnetic moment. 
Open field lines cannot convert Alfv\'en waves into flares/fast waves~\citep{Yuan:2020eor}, meaning that Alfv\'en waves will continue to propagate outward without dissipating. 
We can spot this as an $E^\phi\sim 0$ region in the wave shell corresponding to the north and south pole of the magnetic moment, where field lines are open.

While the overall situation is very similar in the case of a neutron star binary (bottom row), there are subtle differences. 
Since the anti-aligned configuration we show in Fig.~\ref{fig:NSBH_aligned} features connected flux tubes, counter-propagating Alfv\'en waves can be launched along those flux tubes, where they will intersect. 
This requires the presence of a secondary neutron star and does not happen for the mixed binary case. According to Eq.~\eqref{eqn:d_elsasser}, such waves will interact, as shown in the polar region above the binary. 
In principle, interacting Alfv\'en waves can trigger a turbulent cascade via three wave interactions~\citep{Goldreich:1994zz}. 
Numerical studies indicate, however, that many interactions ($>100$ crossings) are required before a turbulent cascade sets in~\citep{Ripperda:2021pzt}, which we do not observe here (our simulations feature at most 1-2 crossings). 
Instead, conversion of Alfv\'en waves into fast waves~\citep{Yuan:2020eor,Chen:2022yci,Chen:2024jvx} is observed in the inner region of both systems, as denoted by the appearance of large amplitude wave-like patterns in $E^\phi$.
 These could provide additional shock-powered dissipation through monster shocks~\citep{Chen:2022yci,Beloborodov:2023lxl,Most:2024qgc}.

Secondary fast waves along the equator are also present (just as in the mixed binary case), but are suppressed compared to the large scale background twist dynamics of the emerging connected flux tube (large positive $E^\phi$ region in Fig.~\ref{fig:NSBH_aligned}.
This is a result of orbital flaring, which for realistic orbital separations will however be strongly suppressed compared to the Alfv\'en wave dynamics \citep{Most:2020ami,Most:2022ojl}.
In either case, there is clear production of a flare with trailing current sheets in the equatorial plane, just as the mixed binary.
This different angular shape of the flares leads us to conclude that at least for misaligned binary neutron star systems, there will be an angular dependence on the emission of resonant shattering flares.

\begin{figure*}
    \centering
\includegraphics[width=\textwidth]{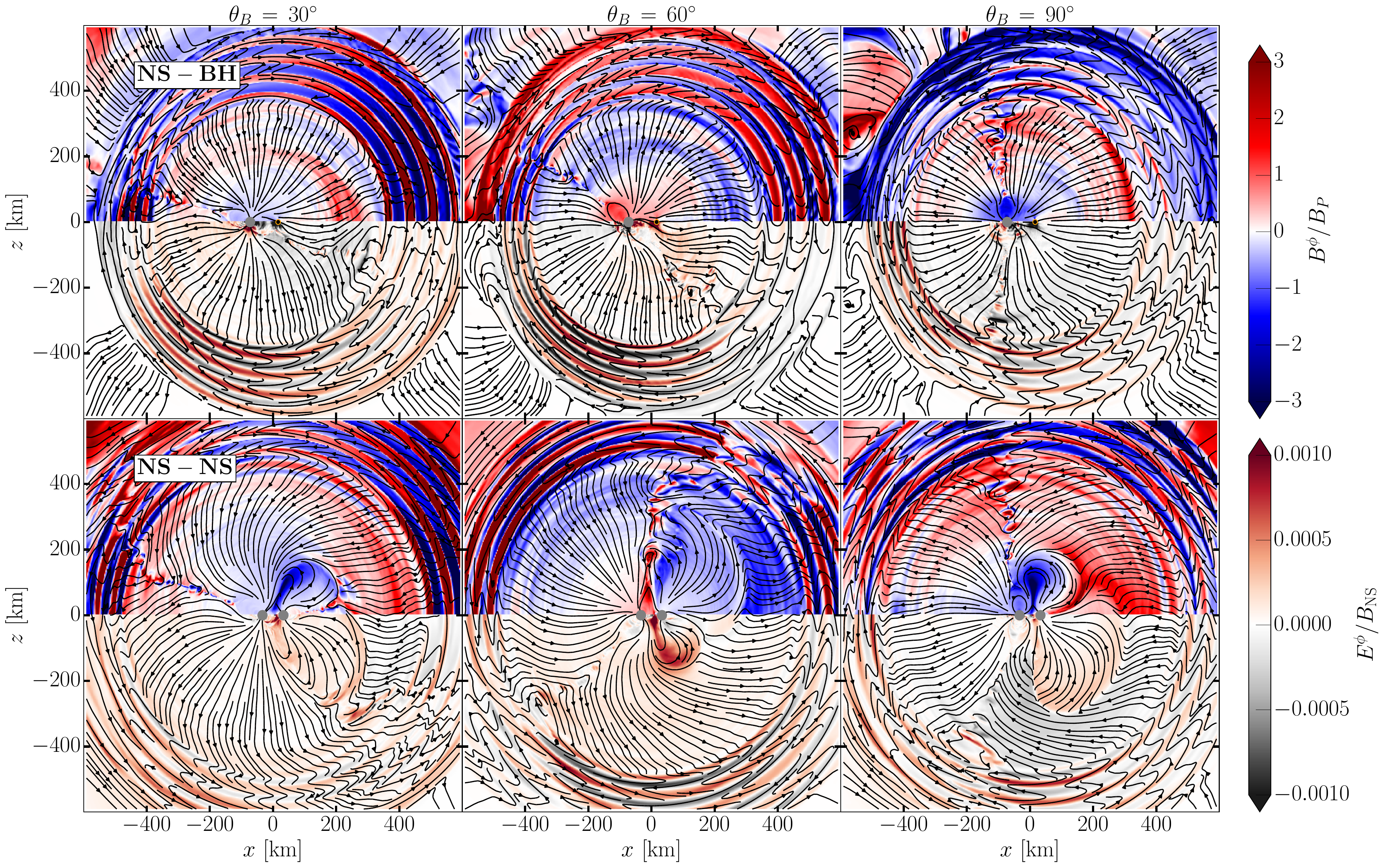}
    \caption{Same as Fig.~\ref{fig:NSBH_aligned} but showing the final times $t-t_{\rm launch}=2.2\,$ms for different initial magnetic field inclinations, $\theta_B$. We do not repeat the $\theta_B=0^\circ$ case, as it is overall similar to $\theta_B=30^\circ$. In all cases, Alfv\'en waves drive flares detaching from the compact binary magnetosphere.}
    \label{fig:inclination}
\end{figure*}

\subsection{Dependence on the  magnetic field geometry}\label{sec:results_parameters}

The alignment of the magnetic field prior to merger is not known, although dipole-dipole interactions may drive the system to near (anti-) alignment~\citep{Aykroyd:2023cvt,Lander:2018und}.
Since tidal excitations have a preferred direction relative to the orbital axis, nonradial shearing motion induced by tidal forces will likely be parallel to the orbital plane. 
Misalignment between the shearing motion and the magnetic axis (akin to a misalignment between the spin and the magnetic moment in a pulsar) will affect flaring emission, since only Alfv\'en waves launched on closed field lines will cause a flaring transient.
One would then expect a reduction in the flare power as a function of the magnetic inclination angle, $\theta_B$. 
Accordingly, we have performed several simulations investigating the impact of the inclination angle $\theta_B$, shown in Fig.~\ref{fig:inclination}.

Starting out with mixed binaries (top panels), Alfv\'en waves in partially inclined configurations, $\theta_B = 30^\circ, 60^\circ$ (left and middle panels), cause similar flaring to aligned configurations, $\theta_B=0^\circ$ (top right panel of Fig.~\ref{fig:NSBH_aligned}). 
The main differences are in the orientation of the transient current sheet (visible via plasmoids formed in the tearing unstable sheet) and the ejection angle of the main plasmoid, which is approximately perpendicular to the magnetic field axis. 
Due to the increased twist (likely of shorter field lines), the emission of secondary fast waves behind the main Alfv\'en wave train is enhanced. 
For $\theta_B=90^\circ$, the situation is different. 
Because now the toroidal perturbation axis and the magnetic moment are misaligned, most Alfv\'en waves are launched on open field lines. 
These Alfv\'en waves steepen fully to nonlinearity and can be seen as large amplitude waves distorting the field lines but they do not dissipate.

For binary neutron star systems (bottom panels) and motivated by the results of previous inclination surveys for compact binary magnetospheres~\citep{Most:2022ojl,Ponce:2014hha}, we consider the case where one of the neutron stars has a magnetic field anti-aligned with the orbital axis, whereas for the other neutron star we vary the magnetic moment inclination.
The main difference in the background field is that rather than having long connected flux tubes extending to high latitudes, misaligned systems feature complicated twist geometries, where the twisted flux tubes extend largely in the angular rather than the radial direction~\citep{Cherkis:2021vto}. 
In this sense, a small fraction of Alfv\'en waves can become trapped, never reaching sufficiently large radii to become non-linear. 
These trapped Alfv\'en waves may ultimately convert to fast waves, albeit at a lower efficiency than the main dynamics discussed here \citep{Yuan:2020eor,Bernardi:2024upq}.

For larger inclination angles, $\theta_B=60^\circ$, similar dynamics sets in.
Fast waves and flares are being formed in both hemispheres, but with a stronger inclination angle. 
In the upper hemisphere, the complicated dynamics of the interacting field lead to an inflated fluxtube connecting both stars, ultimately detaching and causing an orbital-motion powered flare, which is almost orthogonal to the one powered by nonlinear Alfv\'en waves.

Overall, the inclination angle of the dipole magnetic filed will introduce anisotropies in subsequent emission. 
In the following, we quantify the corresponding energetics of the flares.

\subsection{Flare properties}\label{sec:results_flares}

\begin{figure}
    \vskip 1cm
    \centering
\includegraphics[width=0.48\textwidth]{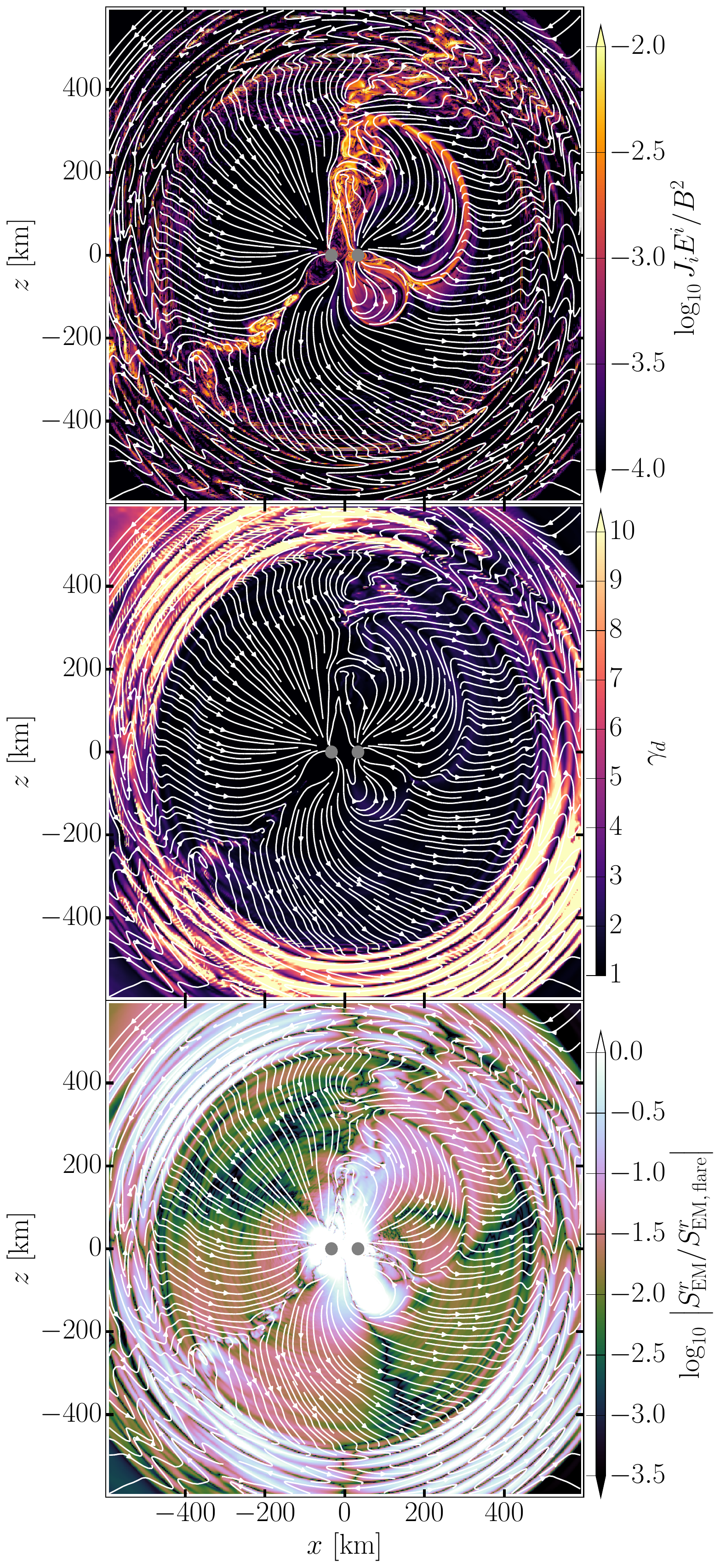}
    \caption{Late stage compact magnetosphere evolution for a binary neutron star system with magnetic field inclination angle $\theta_B=60^\circ$ in the meridional plane. Shown in color are the current sheets, in terms of the dissipative current $J^i$, the drift Lorentz factor $\gamma_d$, as well as the radial Poynting flux, $S_{\rm EM}^r$ normalized to its maximum value, $S_{\rm EM\,, flare}$, in the outgoing flare.}
    \label{fig:diss}
\end{figure}

\begin{figure}
    \centering
    \includegraphics[width=0.48\textwidth]{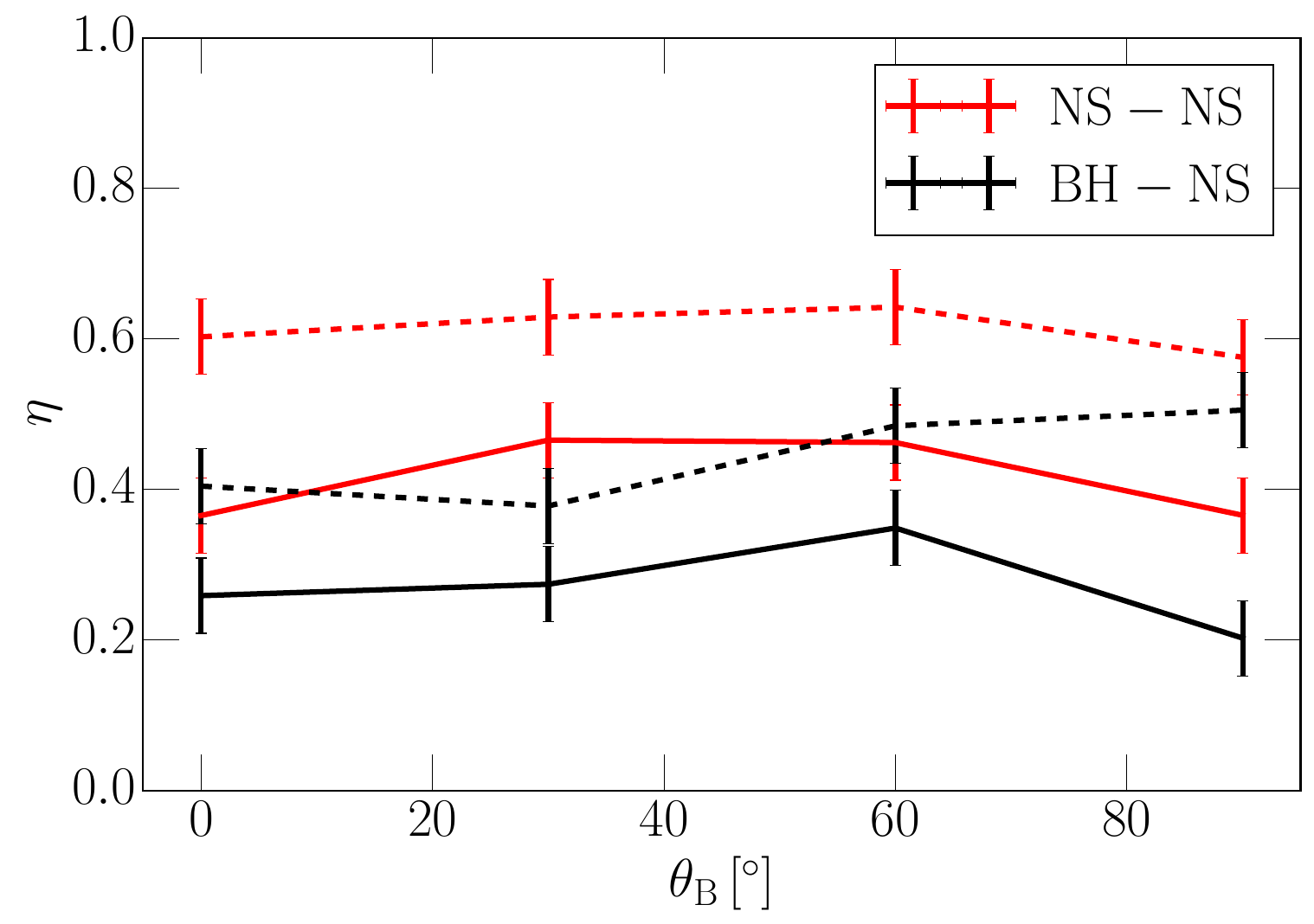}
    \caption{Conversion efficiency $\eta$ of the initial Alfv\'en wave into the final flare. Shown are results for black hole (BH) -- neutron star (NS) and NS-NS systems. Solid lines correspond to waves with drift Lorentz factor $\gamma_d > 5$, which approximately filters out the fast wave component, whereas dashed lines correspond to the entire energy in the final outflow. Error bars quantify the uncertainty in choosing the cut-off on $\gamma_d$.}
    \label{fig:conv_eff}
\end{figure}

We finally provide a quantitative analysis of the flare properties. 
To this end, we focus on a representative case for a neutron star binary systems with $\theta_B=60^\circ$, for which we show the final flaring state in Fig.~\ref{fig:diss}. 
We are mainly interested in the available dissipation channels. 
In that regard, the properties of the flare as well as the energy conversion is very similar to results for single star magnetospheres~\citep{Yuan:2020ayo,Bernardi:2024upq,Mahlmann:2024gui}.

First, we consider reconnection. 
Reconnection in pulsar current sheets can power high energy transients~\citep{Uzdensky:2012tf}. 
In binary magnetospheres this dissipation channel contributes about $10\%$ of the overall Poynting flux~\citep{Most:2020ami,Carrasco:2020sxg,Carrasco:2021jja,Most:2023unc}, which is in line with Alfv\'en-wave dynamics in single neutron star magnetospheres~\citep{Yuan:2020ayo, Yuan:2022uqt}.

We show the local dissipation rate $J_i E^i$ in the top panel.
Nonlinearity of the Alfv\'en waves has created a strong perturbation of the orbital current sheet, which fragments into plasmoids~\citep{Uzdensky:2012tf}. 
In principle, mergers of these plasmoids can produce radio nano-shot emission~\citep{Philippov:2019qud,Lyubarsky:2018vrk} provided that the magnetic field strength is on the order of $10^8\, \rm G$ in the current sheet. 
A similar behavior has been found for flares caused by orbital motion~\citep{Most:2020ami}. Energetically, this emission channel will likely be subdominant.

We similarly find that in the core of the large plasmoid of the Alfv\'en-wave-driven flare substantial dissipation is present. 
This has been argued by \citet{Yuan:2020ayo} to be a source of X-ray emission. 
Since our prescription for dissipation is ad-hoc (using an artificial parallel conductivity), we cannot meaningfully quantify this dissipation beyond the upper percentage limit quoted above.

Second, and beyond dissipation in current sheets, we also quantify the properties of the converted flare directly. 
To do so, we compute the drift Lorentz factor, $\gamma_d$, which is very large $\gamma_d \gg 10$ (middle panel). 
Indeed, the flare will continue to expand, accelerate and become a flat pancake resembling a blast wave with $\gamma_d \gg 100$ at distances of $10^{13}\, \rm cm$~\citep{Yuan:2020ayo}.
Portions of the initial Alfv\'en wave on open field lines become nonlinear but have low drift Lorentz factor (upper right corner of the center panel in Fig. \ref{fig:diss}). This feature is common among all of our simulations.

Finally, we analyze the energetics of the flare. 
Since our simulations do not model the crust itself, we cannot correlate the energy contained in the flare with that in the crust directly. 
However, we can compute how much of the energy we initially inject in the Alfv\'en wave ends up in the flare by computing the energy as a volume integral,
\begin{align}
    E_{\mathcal{S}} = \int_{\mathcal{S}}\, {\rm d}V\, e_{\rm EM} = \int_{\mathcal{S}}\, {\rm d}V \frac{1}{2} \left(E^2 + B^2\right)\,,
\end{align}
contained in a spherical shell $\mathcal{S}$. 
Here $e_{\rm EM} = \frac{1}{2}\left(E^2 + B^2\right)$ is the electromagnetic energy density. 
Choosing $\mathcal{S}$ to contain the initial Alfv\'en wave/final flare we obtain the energies $E_{\rm Alfv\acute{e}n}$ and $E_{\rm Flare}$, respectively. 
The latter time is chosen to correspond to a propagation distance of at least $500\, \rm km$ from the binary when the Alfv\'en wave has become fully nonlinear. 
The resulting energy conversion efficiency, $\eta$, is given in Fig.~\ref{fig:conv_eff}. 
Regardless of dipole inclination the efficiency is always $\simeq 20\%$ for black hole -- neutron star and $\simeq 40\%$ for binary neutron star systems when considering only the part of the outflow that attains high Lorentz factor ($\gamma_d >5$), and $\simeq 40\% -60\%$ in total. 
Since disentangling the energy content of Alfv\'en and fast waves is already complicated in axisymmetric spacetimes~\citep{Bernardi:2024upq,Yuan:2020eor,Mahlmann:2024gui}, we use this distinction as a simple proxy for the energy that ends up in the flare. 
Overall, these numbers are consistent with findings for single star magnetospheres~\citep{Yuan:2020ayo}.

In addition to the total luminosity, we also comment on the spatial distribution of the electromagnetic energy (Poynting) flux~\citep{Gourgoulhon:2012ffd},
\begin{align}
    S^i_{\rm EM} = \varepsilon^{ijk} E_j B_k - \beta^i e_{\rm EM}\,,
\end{align}
where $\beta^i$ is the coordinate shift vector. This is shown in the lower panel of Fig.~\ref{fig:diss}. 
Here, most of the Poynting flux is carried by the steepened blast wave. 
However, in the upstream of this blast wave the Poynting flux is only 1-2 orders smaller. 
This implies that the local background wind is enhanced substantially over its background state, which in general will be even lower, see Eq.~\eqref{eqn:orbiting_dipole}. 
While this does not have implications for a single Alfv\'en wave event as we show here, it may have implications for multiple staggered Alfv\'en waves as are expected to be launched in a crustal shattering event~\citep{Tsang:2013mca,Neill:2020szr}. 
Indeed, \citet{Yuan:2020ayo} found that this enhancement may strongly increase the feasibility for shock-maser powered radio emission~\citep{Metzger:2019una,Beloborodov:2019wex}. 
This finding critically motivates discussion on potential radio transients in Sec.~\ref{sec:picture}.

\section{Conclusions}
\label{sec:conclusions}

Tidal interactions in coalescing binaries involving neutron stars can resonantly excite modes at the crust-core interface~\citep{Penner:2011br,Tsang:2011ad} that shatter the neutron star crust~\citep{Tsang:2013mca} and launch waves into the compact binary magnetosphere prior to merger. 
Since relevant modes will largely be nonradial~\citep{Suvorov:2022ldw,Sagert:2022gwu}, such waves should predominantly be Alfv\'en waves that can nonlinearly steepen in the background of the closed zone~\citep{1989ApJ...343..839B}. 
In the nonlinear phase, the Alfv\'en waves can convert into a blast wave/flare~\citep{Yuan:2020ayo,Yuan:2022uqt}, which can then power electromagnetic transients.

In this work, we provided an investigation of nonlinear Alfv\'en-wave dynamics in compact binary magnetospheres using general-relativistic force-free electrodynamics simulations. 
The computational need to capture the nonlinear phase of Alfv\'en-wave dynamics required simulations on hundreds of GPUs.
Since our simulations did not model the dynamics of the crust, we have injected a monochromatic wave package and tracked its evolution through the fully nonlinear phase.
In reality, resonances in the crust \citep{2020ApJ...897..173B} and fracturing patterns on the surface \citep{Thompson:2016dkd}, will likely cause a variety of different oscillation frequencies to be injected into the magnetosphere. This could affect the potential intersection of waves (when realistic magnetizations are included), the presence or onset of Alfvenic turbulence \citep{Goldreich:1994zz,TenBarge:2021qmk,Ripperda:2021pzt}, and the power contained in the resulting flares. Consistent models for the shattering of the neutron star crust will be needed to answer these questions. 

Our results demonstrate the formation of a high velocity flare in binary neutron star and black hole -- neutron star systems irrespective of the inclination of the magnetic field.

Overall, at least $20\%-40\%$ of the energy initially injected gets converted into the final flare, slightly less than for single star magnetospheres~\citep{Yuan:2020ayo}. 
Additionally, the flare enhanced the wind in its upstream, thus increasing the prospect for shock-powered radio emission in a multi-flare scenario, see Sec.~\ref{sec:picture}. 
In line with previous work on compact binary magnetospheres~\citep{Carrasco:2020sxg,Most:2020ami,Carrasco:2021jja,Most:2022ojl,Most:2023unc}, the system features copious reconnection-mediated dissipation channels which can power secondary X-ray emission.

Our results represent an important step toward understanding the feasibility of the crustal shattering scenario~\citep{1989ApJ...343..839B,Tsang:2011ad,Penner:2011br}.
We demonstrate numerically that the magnetospheric dynamics indeed give rise to flare launching based on local shearing motion on the neutron star surface. 
Equipped with such a framework, it is possible to study the propagation and dynamics of realistic crustal shattering or star quake events~\citep{2020ApJ...897..173B,Sagert:2022gwu}, potentially enabling more direct connections between astrophysical observables and the nuclear physics of the crust~\citep{Neill:2020szr,Neill:2022psd}.

\section*{Acknowledgments}
The authors are grateful for insightful discussions with A. Beloborodov, O. Blaes, A. Bransgrove, D. Neill, A. Philippov, E.S. Phinney, J. Read, B. Ripperda, L. Sironi, A. Suvorov, C. Thompson, Y. Yuan, and B. Zhang.
ERM acknowledges support from the National Science Foundation under Grant No. AST-2307394.
IL and KC were supported by a grant from the Simons Foundation (MP-SCMPS-00001470).
The simulations were performed on DOE NERSC supercomputer Perlmutter under grant m4575. Additional simulations were performed on DOE OLCF Summit under allocation AST198.
Preliminary work was also performed on NSF Frontera supercomputer under grant AST21006. 
ERM further acknowledges the use of Delta at the National Center for Supercomputing Applications (NCSA) through allocation PHY210074 from the Advanced
Cyberinfrastructure Coordination Ecosystem: Services \& Support (ACCESS)
program, which is supported by National Science Foundation grants
\#2138259, \#2138286, \#2138307, \#2137603, and \#2138296.

\software{AMReX \citep{amrex},
	  FUKA \citep{Papenfort:2021hod},
	  Kadath \citep{Grandclement:2009ju},
	  matplotlib \citep{Hunter:2007},
	  numpy \citep{harris2020array},
	  scipy \citep{2020SciPy-NMeth}
}

\bibliography{inspire,non_inspire}

\end{document}